\documentclass[aps,prc,reprint,groupedaddress]{revtex4-1}
\usepackage{epsfig}
\usepackage{multirow}
\begin{document}


\title{Structure effects in the $^{15}$N($n,\gamma$)$^{16}$N radiative capture reaction from the Coulomb dissociation of $^{16}$N}


\author{Neelam}
\affiliation{Department of Physics, Indian Institute of Technology - Roorkee, 247667,
INDIA}
\author{Shubhchintak}
\thanks{Present address : Dept. of Physics and Astronomy, Texas A$\&$M University - Commerce, Commerce, TX-75428, USA}
\author{R. Chatterjee}
\email{Email address: rcfphfph@iitr.ac.in}
\affiliation{Department of Physics, Indian Institute of Technology - Roorkee, 247667,
INDIA}


\date{\today}

\begin{abstract}
\begin{description}
\item[Background]
The $^{15}$N($n, \gamma$)$^{16}$N reaction plays an important role in red giant stars and also in inhomogeneous big bang nucleosynthesis. However, there are controversies regarding spectroscopic factors of the four low-lying states of $^{16}$N, which have direct bearing on the total direct capture cross section and also on the reaction rate. Direct measurements of the capture cross section at low energies are scarce and is available only at three energies below 500 keV.
\item[Purpose]
The aim of this paper is to calculate the $^{15}$N($n, \gamma$)$^{16}$N radiative capture cross section and its subsequent reaction rate by an indirect method and in that process investigate the effects of spectroscopic factors of different levels of $^{16}$N to the cross section.
\item[Method]
A fully quantum mechanical Coulomb breakup theory under the aegis of post-form distorted wave Born approximation is used
to calculate the Coulomb breakup of $^{16}$N on Pb at 100 MeV/u. This is then related to the photodisintegration cross section of $^{16}$N($\gamma, n$)$^{15}$N and subsequently invoking the principle of detailed balance, the $^{15}$N($n, \gamma$)$^{16}$N capture cross section is calculated. 
\item[Results] 
The non-resonant capture cross section is calculated with spectroscopic factors from the shell model and those extracted (including uncertainties) from two recent experiments. The data seems to favor a more single particle nature for the low-lying states of $^{16}$N. The total neutron capture rate is also calculated by summing up non-resonant and resonant (significant only at temperatures greater than 1 GK) contributions and comparison is made with other charged particle capture rates. In the typical temperature range of $0.1-1.2$ GK, almost all the contribution to the reaction rate comes from capture cross sections below 0.25 MeV.
\item[Conclusion]
We have attempted to resolve the discrepancy in the spectroscopic factors of low-lying $^{16}$N levels and conclude that it would certainly be useful to perform a Coulomb dissociation experiment to find the low energy capture cross section for the reaction, especially below 0.25 MeV. 
 
\end{description}
\end{abstract}




\maketitle

\section {Introduction}
Radiative capture reactions play an important role in stellar nucleosynthesis. At temperatures relevant to these events the corresponding relative energies between the participating nuclei are mainly in the sub MeV scale. Direct measurements of reaction cross section at these low energies are often very difficult. In fact, for the $^{15}$N($n,\gamma$)$^{16}$N reaction direct measurements have been possible only at few energies below 500 keV \cite{meissner}. The situation can be addressed by an indirect method like the Coulomb dissociation method, wherein capture cross sections at low energies can be obtained from Coulomb dissociation measurements at higher energies. The $^{14}$C($n,\gamma$)$^{15}$C \cite{nakamura} and $^{7}$Li($n,\gamma$)$^{8}$Li \cite{horvath} neutron capture reactions are two recent examples where the Coulomb dissociation method have been used to calculate the corresponding radiative capture cross section. 
 Therefore it would be interesting to investigate if in the case of $^{15}$N($n,\gamma$)$^{16}$N too indirect measurements add to a better understanding of capture cross sections at low energies.

The radiative neutron capture $^{15}$N($n, \gamma$)$^{16}$N reaction, plays an important role in the synthesis of heavy elements by $s$-process nucleosynthesis in red giant stars and also in the inhomogeneous Big Bang model \cite{kajino,kawano,dubo}.
This reaction is also a part of the neutron induced chain which leads to the breakout from the CNO cycle and hence depletion of CNO abundances \cite{Wiescher}.  
Being the competing reaction with $^{15}$N($\alpha, \gamma$)$^{19}$F, the neutron capture $^{15}$N($n, \gamma$)$^{16}$N reaction is also important in determining the abundance of fluorine \cite{bardayan,guo}. Furthermore, it is also thought to compete with charged particle capture reactions on $^{15}$N \cite{kawano} and therefore can affect the abundance of heavier mass elements.

The only direct measurement of the $^{15}$N($n, \gamma$)$^{16}$N capture cross section has been performed at neutron lab energies of 25, 152 and 370 keV by Meissner {\it et al.} \cite{meissner}. The direct capture calculations performed in order to explain the data using experimental spectroscopic factor ($C^2S$) \cite{bohne}, show a $p$-wave dominated capture. These $C^2S$ had an inherent uncertainity of 30$\%$. Further, their calculated reaction rates were $30-50{\%}$ smaller than those calculated by Rauscher {\it et al.} \cite{rauscher}. Theoretical calculations by Herndl {\it et al.} \cite{herndl}, in the framework of a hybrid compound and direct capture model ($C^2S$ from Ref. \cite{bohne}) were used to explain the data \cite{meissner} and their calculated rates were in agreement with those of Ref. \cite{rauscher}. Another direct capture calculation has been performed in Ref. \cite{bert} using potential model \cite{radcap} with $C^2S$ from Ref. \cite{bohne}.

In fact, the capture cross section and the rate of the $^{15}$N($n, \gamma$)$^{16}$N reaction strongly depends upon the $C^2S$ of the four low-lying levels (with spin-parity $J^\pi =$ 2$^-$, 0$^-$, 3$^-$ and 1$^-$) in $^{16}$N. The calculations of Refs. \cite{meissner,bert}, could account for the data only when the suggested 30{\%} uncertainty in the $C^2S$ from Ref. \cite{bohne} were considered. But this was not the case with the calculations of Ref. \cite{herndl}. A point also to be noted is that the experimentally extracted $C^2S$ in Ref. \cite{bohne} are almost half as those calculated from shell model \cite{meissner,bohne}, which gives a pure single particle picture of these levels. 
In this regard, an experiment was performed by Bardayan {\it et al.} \cite{bardayan}, where they extracted the $C^2S$ for all these four levels, from the measured angular distribution of $^{15}$N($d, p$)$^{16}$N. These $C^2S$ values obtained were close to unity and were in good agreement with those suggested by the shell model \cite{meissner,bohne}. However, the $C^2S$ values extracted in a recent experiment \cite{guo} from the measured angular distribution of $^{15}$N($^7$Li, $^6$Li)$^{16}$N, are not in full agreement with either of the previous experiments \cite{bardayan,bohne}. Their $C^2S$ values suggest that the two levels of $^{16}$N (with J$^\pi$ = 2$^-$ and 3$^-$) are good single-particle levels whereas, the other two (with J$^\pi$ = 0$^-$ and 1$^-$) are not.

With this background we present an indirect method of calculating the $^{15}$N($n, \gamma$)$^{16}$N radiative capture cross section from the Coulomb breakup of $^{16}$N on Pb at 100 MeV/u beam energy. The Coulomb breakup theory is fully quantum mechanical and is calculated under the post-form finite range distorted wave Born approximation (FRDWBA) \cite{rc}. The theory is mainly analytic in nature given that pure Coulomb wave functions are used in the calculation and that the dynamics can be exactly evaluated. Thus, the main aim of this paper is to use this theory to calculate the $^{15}$N($n, \gamma$)$^{16}$N radiative capture cross section and its subsequent reaction rate by an indirect method and in that process investigate the effects of $C^2S$ of different levels of $^{16}$N to the cross section. Previously this theory has been successfully applied to calculate the radiative neutron capture cross sections and subsequent rates of the reactions $^8$Li$(n,\gamma)^9$Li \cite{pb1} and $^{14}$C$(n,\gamma)^{15}$C \cite{sc} from the Coulomb breakup of $^9$Li and $^{15}$C, respectively.

The paper is organized in the following way. Section \ref{sec:2}, contains a brief formalism of the Coulomb breakup process and the capture cross section. In section \ref{sec:3}, we present our results, where we discuss the capture cross section and rate of the $^{15}$N($n,\gamma$)$^{16}$N reaction from the Coulomb dissociation of $^{16}$N and finally in section \ref{sec:4} we present our conclusions.

\section{Formalism}
\label{sec:2}
\begin{figure}[h]
\centering
\includegraphics[height=5.2cm, clip,width=0.4\textwidth]{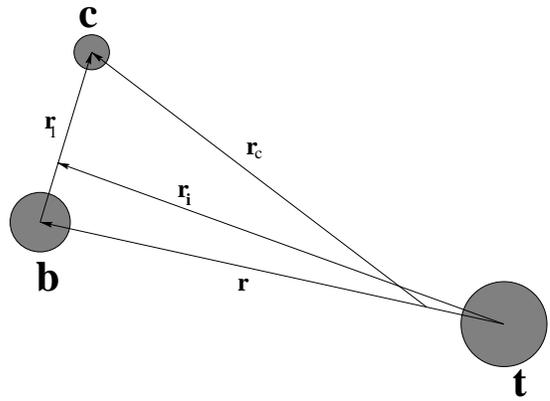}
\caption{\label{fig:1} The three-body Jacobi coordinate system.}
\end{figure}

We consider the elastic breakup of a two-body composite projectile $a$ in the Coulomb field of target $t$ as: $a+t\rightarrow b+c+t$, where the projectile $a$ breaks up into fragments $b$ (charged) and $c$ (uncharged). The three-body Jacobi coordinate system adopted is shown in Fig.~\ref{fig:1}.

The position vectors ${\bf {r}}_{1}$, ${\bf {r}}_{i}$, ${\bf{r}}_{c}$ and ${\bf{r}}$ satisfy the following relations,
\begin{eqnarray}
{\bf{r}}={\bf{r}}_{i}-\alpha{\bf{r}}_{1} ;\hspace{0.2in} {\bf{r}}_{c}=\gamma{\bf{r}}_{1}+\delta{\bf{r}}_{i}\label{a1}.\nonumber    
\end{eqnarray}
$\alpha$, $\gamma$ and $\delta$ are the mass factors, given by:
\begin{eqnarray}
\alpha =\frac{{m_{c}}}{m_{c}+m_{b}};\hspace{0.2in}  \delta =\frac{m_{t}}{m_{b}+m_{t}};\hspace{0.2in} \gamma=(1-\alpha\delta),\label{a2}\nonumber
\end{eqnarray}
where $m_{b}$, $m_{c}$ and $m_{t}$ are the masses of fragments $b$, $c$ and $t$, respectively.

The triple differential cross section for the reaction in terms of relative coordinates is given by
\begin{eqnarray}
\frac{d^3\sigma}{dE_{bc}d\Omega_{bc}d\Omega_{at}} &=& \frac{2\pi}{\hbar v_{at}}\frac{\mu_{bc}\mu_{at}p_{bc}p_{at}}{h^6}\nonumber\\
&\times& \sum_{\ell m}\frac{1}{(2\ell+1)}|\beta_{\ell m}|^2,\label{a4.1}
\end{eqnarray}
where $v_{at}$ is the $a-t$ relative velocity in the entrance channel and $E_{bc}$ is the $b-c$ relative energy in the final channel.
 $\mu_{bc}$ and $\mu_{at}$ are the reduced masses, $\Omega_{bc}$ and $\Omega_{at}$ are solid angles and $p_{bc}$ and $p_{at}$ are appropriate linear momenta corresponding to the $b-c$ and $a-t$ systems, respectively. $\beta_{\ell m}$ is the reduced amplitude in post form FRDWBA, given by
\begin{eqnarray}
\beta_{\ell m} &=& \left\langle e^{i(\gamma{\bf q_c} - \alpha{\bf K}).{\bf r_1}}\left|V_{bc}\right|\phi_{a}^{\ell m}({\bf r}_1)\right\rangle \nonumber\\
&\times&\left\langle \chi_{b}^{(-)}({\bf q_b},{\bf r_i})e^{i\delta {\bf q_c}.{\bf r_i}}|\chi_{a}^{(+)}({\bf q}_a,{\bf r_i})\right\rangle. \label{a4.2}
\end{eqnarray}
The first term containing the projectile bound state wave function $\phi_{a}^{\ell m}({\bf r}_1)$ of any angular momentum $\ell$ (with projection $m$) is the structure part, while the second term containing the Coulomb distorted waves $\chi^{(\pm)}$ describes the dynamics of the reaction and further can be expressed analytically in terms of the bremsstrahlung integral ~\cite{Nordsieck}. $V_{bc}$ is the interaction between $b$ and $c$ in the initial channel. In Eq. (\ref{a4.2}), ${\bf K}$ is an effective local momentum appropriate to the core-target relative system (see appendix A) and ${\bf q}_i$'s ($i = a, b, c$) are the Jacobi wave vectors of the respective particles. For more details on these quantities we refer to  Ref. \cite{rc}.

The relative energy spectra ($\frac{d\sigma}{dE_{bc}}$) of the reaction can be obtained from Eq. (\ref{a4.1}) by integrating over the appropriate solid angles.

Then, the photodisintegration cross section ($\sigma_{\gamma, n}^{\pi\lambda}$) for the reaction $a+\gamma\rightarrow b+c$ can be related to the relative energy spectra as,
\begin{eqnarray}
\sigma_{\gamma, n}^{\pi\lambda} = \frac{E_\gamma}{{\rm n}_{\pi\lambda}}\frac{d\sigma}{dE_{bc}}, \label{a4.5}
\end{eqnarray}
provided a single multipolarity ($\pi\lambda$) dominates. Here $\pi$ stands for electric or magnetic type and $\lambda$ is the multipolarity. 



In Eq. (\ref{a4.5}), $E_{\gamma}=E_{bc}+Q$ is the photon energy with ${\it Q}$ as the ${\it Q}$-value of the reaction and ${\rm n}_{\pi\lambda}$ is the equivalent photon number which depends upon the $a - t$ system \cite{Bertulani}. For more details of the method one is referred to Refs. \cite{sc,Bertulani,Gade,akrm}.



The radiative capture cross section $\sigma_{n, \gamma}$ can then be calculated utilising the principle of detailed balance,
\begin{eqnarray}
\sigma_{n, \gamma} = \frac{2(2j_a+1)}{(2j_b+1)(2j_c+1)}\frac{k_{\gamma}^2}{k^2_{bc}}\sigma_{\gamma, n}^{\pi\lambda}, \label{a4.6}
\end{eqnarray}
where $j_a$, $j_b$ and $j_c$ are the spins of  particles $a$, $b$ and $c$, respectively. $k_\gamma$ and $k_{bc}$ are the wave numbers  of the photon and that of relative motion between $b$ and $c$, respectively.

The non-resonant reaction rate per mole $N_A\langle\sigma v\rangle_{nr}$ ($N_A$ being Avogadro constant)  can be calculated from the neutron capture cross section $\sigma_{n, \gamma}(E_{bc})$ as:
\begin{eqnarray}
N_A\langle\sigma v\rangle_{nr} &=& N_A \sqrt{\frac{8}{(k_BT)^{3} \pi\mu_{bc}}} \nonumber\\
&\times& \int^\infty_0{\sigma_{n, \gamma}(E_{bc})\hspace{0.1cm} E_{bc} \hspace{0.1cm} exp(-\frac{E_{bc}}{k_BT})\hspace{0.1cm}dE_{bc}}, \label{a4.7}
\end{eqnarray}
where $k_B$ is the Boltzmann constant and $T$ is the temperature in Kelvin (K). We shall show subsequently (section \ref{sec:3.3}) that the significant contribution to the reaction rate comes from a very small range of energies and hence the whole integration range in the above equation need not be considered.

In case of narrow resonances, the capture cross section can be obtained by using the Breit-Wigner formula. In such a case the reaction rate per mole can be easily expressed as the sum over individual resonances with energy $E_i$ \cite{iliadis} as
\begin{eqnarray}
N_A\langle\sigma v\rangle_r &=& 1.54 \times 10^{11} (\mu_{bc} T_9)^{-3/2} \nonumber\\
   &\times& \sum_i (\omega\gamma)_i exp\big(\frac{-11.605 E_i}{T_9}\big), \label{a4.8}
\end{eqnarray} 
with $\omega\gamma$ being the resonance strength and $T_9$ is the temperature in units of $10^9$ K. The total rate per mole $N_A\langle\sigma v\rangle$ is then the sum of non-resonant and resonant contributions. 
\section{Results and discussions}
\label{sec:3}
\subsection{Structure of $^{16}$N}
\label{sec:3.1}

$^{16}$N has one-neutron separation energy ($S_n$) of 2.491 MeV in its ground state having $J^\pi$ = 2$^-$. There are three low-lying excited states with $J^\pi =$ 0$^-$, 3$^-$ and 1$^-$ at energies 0.120, 0.298 and 0.397 MeV above the ground state, respectively. Two states 2$^-$ and 3$^-$ are formed by the coupling of 1$d_{5/2}$ $\nu$ with the 1/2$^-$ ground state of $^{15}$N, whereas the other two states 0$^-$ and 1$^-$ are formed by the coupling of 2$s_{1/2}$ $\nu$ with the 1/2$^-$ ground state of $^{15}$N. All these four levels have been suggested to contribute to the direct capture cross section of $^{15}$N$(n,\gamma)^{16}$N \cite{meissner,herndl,bert} and the capture process is dominated by E1 transitions \cite{bert}.
Apart from these, the 862 keV resonance is the only relevant resonance which has been suggested to contribute to the reaction rate at high temperature ($> 1$ GK) \cite{meissner}.

In our study, we calculate the bound state wave function of the projectile (which is the only input in our theory) by assuming a Woods-Saxon interaction between the valence neutron and the charged core. The depth (V$_0$) of the potential is adjusted to reproduce the binding energy. The radius and diffuseness parameters are taken to be 1.25 fm and 0.65 fm, as in Refs. \cite{bert,guo}. However, for the sake of completeness, we have also investigated the effect of changing the radius and diffuseness parameters by $20\%$ on our results in appendix B.

We use shell model $C^2S$ values which considers low-lying $^{16}$N levels as good single-particle states.
In fact, these are also supported by the experiment performed in Ref. \cite{bardayan}. Another support to our choice comes from isospin symmetry, given that the $C^2S$ for low-lying four levels in mirror nucleus $^{16}$F are near unity \cite{lee,stefan}. Nevertheless, we have also performed our calculations with the $C^2S$ of Refs. \cite{bohne,guo} for the sake of completeness.

\begin{table}[ht]
\begin{center}
\caption{Depths (V$_0$) of the Woods-Saxon potential obtained
corresponding to neutron binding energies ($S_n$) of four low-lying states of $^{16}$N. The shell model $C^2S$ (OXBASH) of these levels are from Ref. \cite{meissner}. 
The values of the radius and diffuseness parameters are taken to be 1.25 fm and 0.65 fm, respectively.}
\vspace{0.08cm}
\begin{tabular}{|c|c|c|c|c|}
\hline\hline
$J^{\pi}$                & configuration   & $S_n$  & 
V$_0$  & $C^2S$\\ 
          &  & (MeV)  & (MeV) & \\
\hline 
2$^-$ & $^{15}$N(1/2$^-$)$\otimes$1$d_{5/2}\nu$ & 2.491   & 58.06  & 0.93 \\
0$^-$ & $^{15}$N(1/2$^-$)$\otimes$2$s_{1/2}\nu$ & 2.371 & 53.89 & 0.95\\
3$^-$ & $^{15}$N(1/2$^-$)$\otimes$1$d_{5/2}\nu$ & 2.193 & 45.04 & 0.87\\
1$^-$ & $^{15}$N(1/2$^-$)$\otimes$2$s_{1/2}\nu$ & 2.094 & 49.38 & 0.96\\
\hline
\hline
\end{tabular}
\end{center}
\end{table}

In Table I, we show the respective depths of the Woods-Saxon potential obtained corresponding to neutron removal from all four levels mentioned above, along with their $S_n$ and $C^2S$ values. 

\subsection{Capture cross section}
\label{sec:3.2}

The first step to calculate the capture cross section is to calculate the relative energy spectra, which we have done for the Coulomb breakup of $^{16}$N on a Pb target at 100 MeV/u beam energy for all projectile bound state configurations mentioned in Table I.
From the relative energy spectra we calculate the photodisintegration cross section for the reaction $^{16}$N$(\gamma, n)^{15}$N, using Eq. (\ref{a4.5}). This is the key step in using Coulomb dissociation as an indirect method in nuclear astrophysics. 
Furthermore, given that the gamma ray transition corresponding to all four levels of $^{16}$N of present interest are all dominated by E1 multipolarity \cite{bert}, so the conditions for the applicability of Eq. (\ref{a4.5}) are fulfilled.
The photodisintegration cross sections are then used to calculate the radiative capture cross sections by applying the principle of detailed balance [Eq. (\ref{a4.6})].
\begin{figure}[h]
\centering
\includegraphics[height=8.5cm, clip,width=0.45\textwidth]{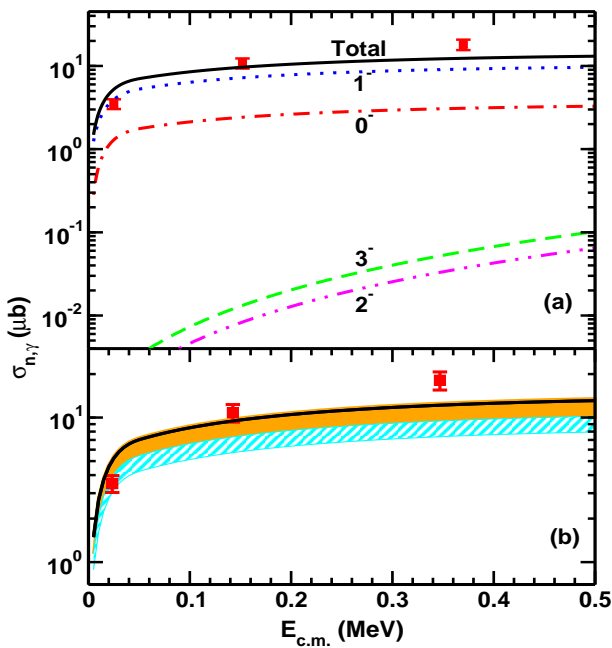}
\caption{\label{fig:2} (Color online) (a) Total non-resonant $^{15}$N$(n,\gamma)^{16}$N cross section (solid line) obtained by summing up contributions of capture to all four states of $^{16}$N (given in Table I) using their respective shell model $C^2S$. (b) Total non-resonant capture cross section obtained by using the experimentally extracted $C^2S$ (including uncertainties) from Ref. \cite{bardayan} (filled band) and Ref. \cite{guo} (filled pattern) compared with the total non-resonant cross section (solid line) shown in (a). The experimental data in both panels are from \cite{meissner}.}
\end{figure}

In Fig. \ref{fig:2} (a), we show our $^{15}$N$(n,\gamma)^{16}$N non-resonant capture cross section as a function of the center of mass energy (E$_{c.m.}$) and compare it with the experimental data, which are available at three energies in the range $0-500$ keV. 
The solid line corresponds to total non-resonant capture cross section which is obtained by summing up capture contributions to all the four levels of $^{16}$N using their respective shell model $C^2S$ values (given in Table I).

It is clear that among all these four states, nearly all the contribution to the total cross section, come from the 1$^-$ and the 0$^-$ states. Therefore the change in $C^2S$ of these two states can change the total cross section to a significant extent. This point is further elucidated when in Fig. \ref{fig:2} (b), we compare the total capture cross section with experimentally extracted $C^2S$ (including their uncertainties) from Refs. \cite{bardayan} (filled band) and \cite{guo} (filled pattern) with those of the shell model (solid line). Clearly the difference between the $C^2S$ of the 1$^-$ and the 0$^-$ states in these two experiments is the reason for their disagreement among the calculated cross sections in Fig. \ref{fig:2} (b). 
It is clear that with the shell model C$^2S$ (which is also supported by the upper limit of the calculations using C$^2S$ of Ref. \cite{bardayan}), our results are in good agreement with the data. This would support the contention that the low-lying states of $^{16}$N are predominantly single particle in nature.

We also wish to point out that capture cross sections at energies below 500 keV will not have any significance contribution from the 862 keV resonance. However, it could contribute to the reaction rate for temperatures $T_9 > 1$, as will be seen later.

\subsection{Reaction rate}
\label{sec:3.3}
As mentioned earlier, $^{15}$N$(n, \gamma)^{16}$N plays important role in the synthesis of heavier nuclei and also it is considered to compete with other charged particle reactions on $^{15}$N. So it would be interesting to find the rate of the $^{15}$N$(n, \gamma)^{16}$N reaction and compare it with the other charged particle reaction rates.
\begin{figure}[h]
\centering
\includegraphics[height=7.0cm, clip,width=0.45\textwidth]{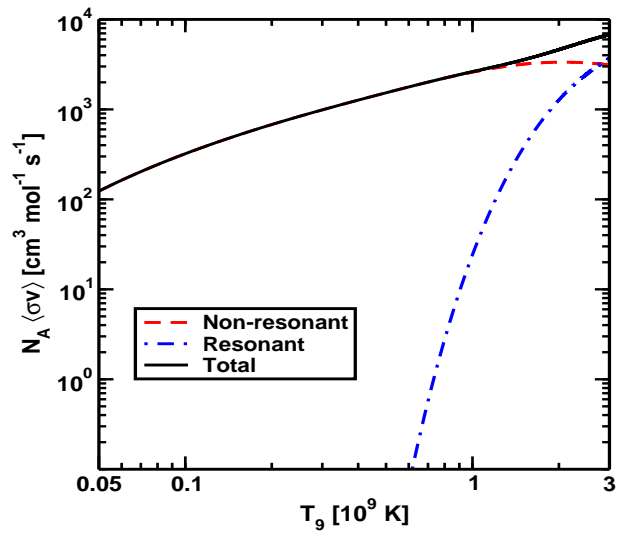}
\caption{\label{fig:3} (Color online) Total  $^{15}$N$(n, \gamma)^{16}$N reaction rate (solid line) obtained by summing up the non-resonant (dashed line) and resonant (dot-dashed line) rates.}
\end{figure}

\begin{figure}[h]
\centering
\includegraphics[height=6.5cm, clip,width=0.45\textwidth]{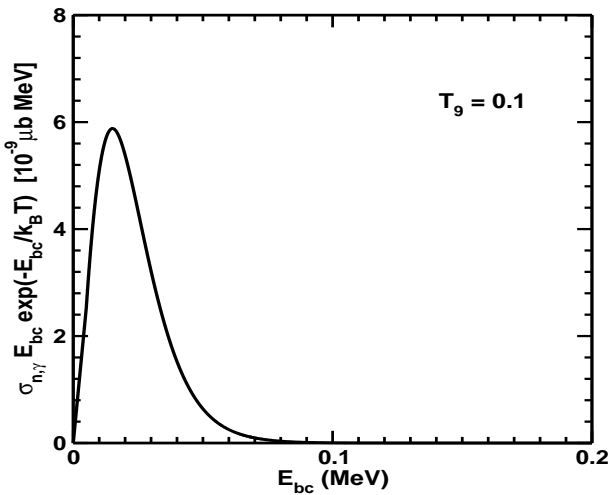}
\caption{\label{fig:4} Integrand in Eq. (\ref{a4.7}), plotted as a function of relative energy (E$_{bc}$) for $T_9$ = 0.1 (typical temperature of AGB stars). For more details see text.}
\end{figure}

In Fig. \ref{fig:3}, we present our $^{15}$N$(n, \gamma)^{16}$N reaction rate in the temperature range $T_9 = 0.05-3$. The total rate (solid line) is the sum of non-resonant (dashed line) and resonant (dot-dashed line) rates. The non-resonant reaction rates are calculated by using Eq. (\ref{a4.7}), where the energy integration has been performed upto $0.5$ MeV, consistent with the energy range shown in Fig. \ref{fig:2}. In order to ensure that we have not missed any contribution to the non-resonant rates at higher energies we plot the integrand in Eq. (\ref{a4.7}) as a function of energy [at $T_9$ = 0.1, a typical temperature of Asymptotic Gaint Branch (AGB) stars] in Fig. \ref{fig:4}. It is clear from the figure that at this temperature almost all the contribution to the non-resonant rate is from the energy range below 0.1 MeV. In fact, we have checked that even at a higher temperature (at $T_9$ = 1) the contribution after 0.25 MeV is negligible. This shows that even in the temperature range relevant for inhomogeneous big bang model i.e. $T_9 = 0.2-1.2$, the maximum contribution to the non-resonant rate of $^{15}$N$(n, \gamma)^{16}$N comes from the energies below 0.25 MeV. Fig. \ref{fig:2} shows that in this energy range our calculated neutron capture cross section are in good agreement with the experimental data.
The resonant rates are calculated using Eq. (\ref{a4.8}) with the parameters given in Ref. \cite{meissner}. They seem to be relevant only for temperatures $T_9 > 1$.

\begin{figure}[h]
\centering
\includegraphics[height=7.0cm, clip,width=0.45\textwidth]{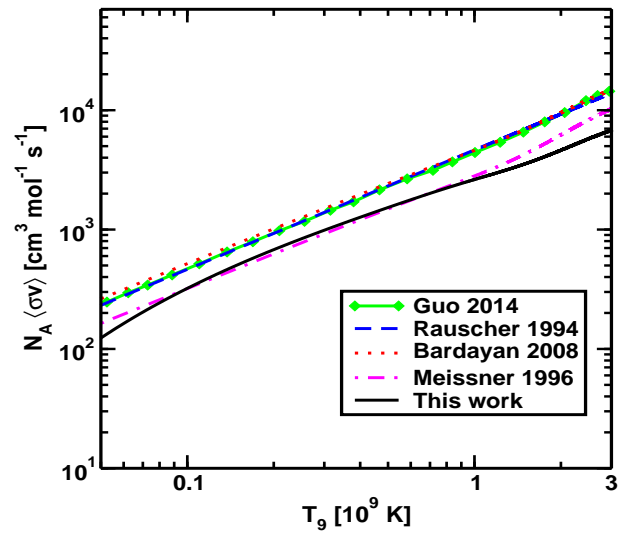}
\caption{\label{fig:5} (Color online) Calculated $^{15}$N$(n, \gamma)^{16}$N reaction rate (solid line) compared with other evaluations based on various experimental estimates of $C^2S$. Guo 2014 : Ref. \cite{guo}; Rauscher 1994 : Ref. \cite{rauscher}; Bardayan 2008 : Ref. \cite{bardayan}, Meissner 1996 : Ref. \cite{meissner}.  For more details see text.}
\end{figure}
As can be expected, different $C^2S$ of the levels of $^{16}$N affects the reaction rate and this has also been seen by several authors so far. 
In Fig. \ref{fig:5}, we compare our total rates (solid line) with the rates from other theoretical predictions and evaluations based on various experimental estimates of $C^2S$ \cite{meissner,bardayan,guo,rauscher}.
The rates reported by Meissner {\it et al.} in Ref. \cite{meissner} (dot-dashed line) are smaller than the rates calculated by Rauscher {\it et al.} \cite{rauscher} (dashed line) by $30-50\%$ and this discrepancy was traced to the different $C^2S$ used. Rates calculated by Bardayan {\it et al.} \cite{bardayan} (dotted line), using their experimentally extracted $C^2S$ were also almost double as compared to those calculated in Ref. \cite{meissner}. The discrepancy of a similar factor has also been reported recently by Guo {\it et al.} in Ref. \cite{guo} (squared line), on comparing their rates with those of Meissner {\it et al.}

It is clear that in the temperature ranges relevant for inhomogeneous big bang model ($T_9 = 0.2-1.2$) and for typical AGB stars ($T_9 \approx 0.1$), our rates are almost same as those in Ref. \cite{meissner}. However, in the same temperature range our predicted rates are slower than those of Refs. \cite{bardayan,guo,rauscher}. We reiterate that our reaction rates are based on capture cross sections derived from a fully quantum mechanical Coulomb breakup theory.

\begin{figure}[h]
\centering
\includegraphics[height=7.0cm, clip,width=0.45\textwidth]{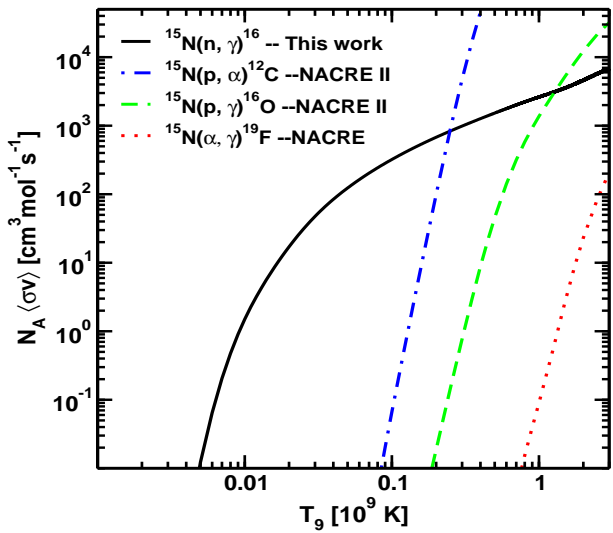}
\caption{\label{fig:6} (Color online) Calculated $^{15}$N$(n, \gamma)^{16}$N reaction rate (solid line) compared with those of $^{15}$N$(p, \alpha)^{12}$C \cite{nacre11} (dot-dashed line), $^{15}$N$(p, \gamma)^{16}$O \cite{nacre11} (dashed line) and $^{15}$N$(\alpha, \gamma)^{19}$F \cite{nacre1} (dotted line).}
\end{figure}

Finally, we turn our attention to the comparison of our rates with those of charged particle capture on $^{15}$N. Fig. \ref{fig:6}, shows the comparison of the rates of reactions $^{15}$N$(n, \gamma)^{16}$N, $^{15}$N$(p, \alpha)^{12}$C, $^{15}$N$(p, \gamma)^{16}$O and $^{15}$N$(\alpha, \gamma)^{19}$F in the temperature range of $T_9 = 0.001-3$. The rates of ($p, \gamma$) and (p, $\alpha$) reactions are from NACRE II compilation \cite{nacre11}, whereas those of ($\alpha$, $\gamma$) are from NACRE compilation \cite{nacre1}. It is clear that at low temperature because of the Coulomb barrier the charged particle capture rates are significantly slower than the ($n,\gamma$) rate. Consequently the $^{15}$N$(n, \gamma)^{16}$N reaction dominates over the $^{15}$N$(p, \alpha)^{12}$C and $^{15}$N$(p, \gamma)^{16}$O reactions in the temperature ranges $T_9$ = $0-0.25$  and $0-1.3$, respectively. Again, given the fact that the rate of the $^{15}$N$(\alpha, \gamma)^{19}$F reaction is very small in the given temperature range, formation of $^{16}$N by neutron capture would be more favourable than the production of $^{19}$F. Therefore, it appears that at temperatures below $T_9 < 0.25$, the probability of consumption of $^{15}$N by neutron capture is more than the proton or alpha capture reactions.

\section{Conclusions}
\label{sec:4}
In summary, we have calculated the $^{15}$N$(n,\gamma)^{16}$N radiative capture cross section and the associated reaction rate by using the Coulomb dissociation of $^{16}$N on Pb at 100 MeV/u, as an indirect method in nuclear astrophysics. Our Coulomb dissociation theory is purely quantum mechanical one, under the aegis of the post-form finite range distorted wave Born approximation. The entire non-resonant continuum is included in the theory and the projectile bound state information is the only input. The local momentum approximation to the transition amplitude allows us to factorize the breakup amplitude into the structure and the dynamics part. This theory has previously been used to study the structure and dynamics of nuclei away from the valley of stability and also to study radiative capture reactions.

We calculate the relative energy spectra in the breakup of $^{16}$N on Pb at 100 MeV/u and calculate the relevant photodisintegration cross sections for the four low-lying states (2$^-$, 0$^-$, 3$^-$ and 1$^-$) of $^{16}$N. The principle of detailed balance is then invoked to calculate the relevant $^{15}$N$(n,\gamma)^{16}$N radiative capture cross sections to the different low-lying states of $^{16}$N. We then bring into focus the state of affairs regarding the spectroscopic factors of these low-lying states. Comparison of our calculations with the available direct capture data \cite{meissner} seems to favour the spectroscopic factors from Ref. \cite{bardayan} (which are similar to the shell model) than those of Refs. \cite{bohne,guo}. 
This would give the credence to the fact that the low-lying levels of $^{16}$N could be single particle in nature. 

With the paucity of direct capture data for this reaction, it would certainly be useful to perform a Coulomb dissociation experiment to find the low energy  capture cross section for the reaction, especially below 0.25 MeV, given that almost all the contribution to the reaction rate comes from this energy range.  An advantage, from an experimental point of view, is that breakup fragments emerge at higher energies (given that the projectile energies are in the range of 100 MeV/u), which in turn facilitates their detection. Moreover, measuring Coulomb dissociation observables like the relative energy spectra and angular distributions one would be able to put constraints on spectroscopic factors. In fact, recently the Coulomb dissociation method has been used to find the neutron capture cross section to different states of $^8$Li \cite{horvath} and also to find the contributions of the projectile excited states in the charged particle capture reactions \cite{fleurot,langer}. 

We also calculate the $^{15}$N$(n,\gamma)^{16}$N reaction rate per mole as a function of temperature. For temperatures relevant for typical AGB stars and for inhomogeneous big bang model, our calculations favor the destruction of $^{15}$N by neutron capture than by proton or alpha capture.

\section*{Acknowledgment}
This text presents results from research supported by the Department of Science and Technology, Govt. of India, (SR/S2/HEP-040/2012). Support from MHRD grants, Govt. of India, to [N] is gratefully acknowledged.



\begin{thebibliography}{99}
\bibitem{meissner} J. Meissner {\it et al.}, Phys. Rev. C {\bf 53}, 977 (1996).
\bibitem{nakamura} T. Nakamura {\it et al.}, Phys. Rev. C {\bf 79}, 035805 (2009). 
\bibitem{horvath} R. Izs{\'a}k {\it et al.}, Phys. Rev. C {\bf 88}, 065808 (2013).
\bibitem{kajino} T. Kajino, G. J. Mathews, and G. M. Fuller, in {\it Heavy-ion physics and astrophysical problems}, edited by S. Kubono (World Scientific, Singapore, 1989).
\bibitem{kawano}  L. H. Kawano, W. A. Fowler, R. W. Kavanagh, R. A. Malaney, Astrophys. J. {\bf 372}, 1 (1991). 
\bibitem{dubo} S. B. Dubovichenko,  Russ. Phys. J., {\bf 56}, 494 (2013).
\bibitem{Wiescher} M. Wiescher, J. G\"orres, H. Schatz, J. Phys. G {\bf 25}, R133 (1999).
\bibitem{bardayan} D. W. Bardayan {\it et al.}, Phys. Rev. C {\bf 78}, 052801(R) (2008). 
\bibitem{guo} B. Guo {\it et al.}, Phys. Rev. C {\bf 89}, 012801(R) (2014).
\bibitem{bohne} W. Bohne {et al.}, Nucl. Phys. A {\bf 196}, 41 (1972).
\bibitem{rauscher} T. Rauscher, J. H. Applegate, J. J. Cowan, F.-K. Thielemann, and M. Wiescher, Astrophys. J. {\bf 429}, 499 (1994).
\bibitem{herndl} H. Herndl, R. Hofinger, J. Jank {\it et al.}, Phys. Rev. C {\bf 60}, 064614 (1999).
\bibitem{bert} J. T. Huang, C. A. Bertulani, V. Guimar\~{a}es, Atomic Data and Nuclear Data Tables {\bf 96}, 824 (2010).
\bibitem{radcap} C.A. Bertulani, Comput. Phys. Commun. {\bf 156}, 123 (2003).
\bibitem{rc} R. Chatterjee, P. Banerjee, R. Shyam, Nucl. Phys. A {\bf 675}, 477 (2000).
\bibitem{pb1} P. Banerjee, R. Chatterjee, R. Shyam Phys. Rev. C {\bf 78}, 035804 (2008).
\bibitem{sc} Shubhchintak, Neelam and R. Chatterjee, Pramana J. Phys. {\bf 83}, 533 (2014).
\bibitem{Nordsieck} A. Nordsieck, Phys. Rev. {\bf 93}, 785 (1954).
\bibitem{Bertulani} C. A. Bertulani, G. Baur, Phys. Rep. {\bf 163}, 299 (1988).
\bibitem{Gade} C. A. Bertulani, A. Gade, Phys. Rep. {\bf 485}, 195 (2010).
\bibitem{akrm} R. E. Tribble, C. A. Bertulani, M. La Cognata, A. M. Mukhamedzhanov and C. Spitaleri, Rep. Prog. Phys. {\bf 77}, 106901 (2014).
\bibitem{iliadis} C. Iliadis, {\it Nuclear Physics of stars}, (Wiley-VCH Verlag GmbH {\&} Co. KGaA, Weinheim, 2007).
\bibitem{lee}  D. W. Lee, K. Per\"aj\"arvi, J. Powell, J. P. O'Neil, D. M. Moltz, V. Z. Goldberg, and J. Cerny, Phys. Rev. C {\bf76}, 024314 (2007).
\bibitem{stefan} I. Stefan {\it et at.}, Phys. Rev. C {\bf 90}, 014307 (2014).
\bibitem{nacre11} Y. Xu, K. Takahashi, S. Goriely, M. Arnould, M. Ohta and H. Utsunomiya, Nucl. Phys. A {\bf 918}, 61 (2013).
\bibitem{nacre1} C. Angulo {\it et al.}, Nucl. Phys. A {\bf 656}, 3 (1999).
\bibitem{fleurot}F. Fleurot {\it et al.}, Phys. Lett. B {\bf 615}, 167 (2005).
\bibitem{langer} C. Langer {\it et al.}, Phys. Rev. C {\bf 89}, 035806 (2014).
\bibitem{lmanag} R. Shyam and M.A. Nagarajan,  Ann. Phys. (N.Y.) {\bf 163}, 265 (1985).
\bibitem{rcprc} R. Chatterjee,  Phys. Rev. C {\bf 68}, 044604 (2003).
\bibitem{zadro} M. Zadro, Phys. Rev. C {\bf 66}, 034603 (2002).


\end{thebibliography}
\appendix
\section*{Appendix A: Validity of the Local momentum approximation}
We use the local momentum approximation \cite{lmanag,rcprc,zadro}, on the
outgoing charged fragment $b$ to obtain the factorization of the breakup amplitude [Eq. (2)]. 
Essentially, it involves the Taylor expansion of $\chi_{b}^{(-)*}({\bf{q}}_{b},{\bf{r}})$ about ${\bf r_i}$, which is exact and helps in separation of the variables ${\bf r_i}$ and ${\bf r_1}$,
\begin{eqnarray}
\chi^{(-)}_b({\bf q}_b,{\bf r}) = e^{-\alpha{\bf r}_1.\nabla_{r_i}}\chi^{(-)}_b({\bf q}_b,{\bf r}_i).\label{f3.7}
\end{eqnarray} 
Then one can approximate the del-operator to an effective local momentum, ${\bf K} (=-i\nabla_{{\bf r}_{i}})$, whose magnitude is given by 
\begin{eqnarray}
{K} = \sqrt{\frac{2\mu_{bt}}{\hbar^{2}}(E_{bt}-V(R))},\label{f3.8}
\end{eqnarray} 
where $\mu_{bt}$ is the reduced mass of the $b-t$ system, $E_{bt}$ is the energy of particle $b$ relative to the target in the c.m. system and $V(R)$ is the  Coulomb potential between $b$ and the target at a distance $R$. Thus, the local momentum {\bf K} is then evaluated at some fixed distance $R$ (10 fm, in our case) and the magnitude is held fixed for all the values of {\bf r}. The direction of {\bf K} is taken to be same as that of the outgoing fragment $b$.
The condition of validity (see eg. \cite{lmanag})
is that the quantity 
\begin{eqnarray}
\eta(r) = \frac{{1\over 2} K(r)}{\Big| \frac{d K(r)}{dr} \Big|}, \label{ea8}
\end{eqnarray}
calculated at some representative distance $R$ should be more than the projectile
radius, $r_a$. 

\begin{figure}[h]
\centering
\includegraphics[height=6.0cm, clip,width=0.45\textwidth]{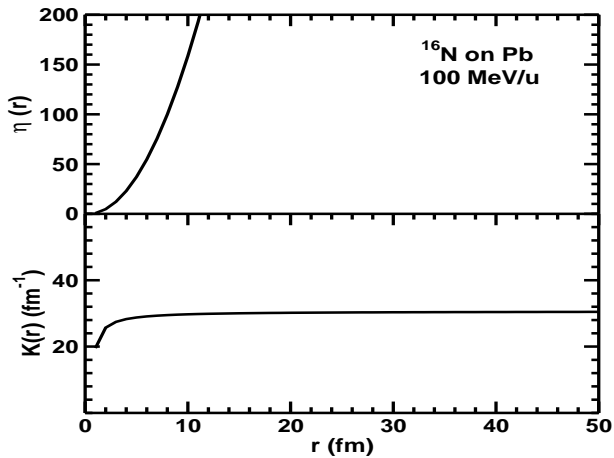}
\caption{\label{fig:7} Variation of $\eta (r)$ (upper half) and ${ K}(r)$
(lower half) with $r$ for the Coulomb breakup of $^{16}$N on Pb in its ground state. For more details, see the text. }
\end{figure}

To check the validity of the approximation, in Fig. \ref{fig:6} we show the variation of
$\eta(r)$ (upper half) and ${ K}(r)$ (the magnitude of the local
momentum) (lower half) as a function of $r$, for the  Coulomb breakup reaction
$^{16}$N + Pb $\rightarrow$ $^{15}$N + $n$ + Pb at the beam energy of  
100 MeV/u. At $r =10$ fm, $\eta(r) >> r_a$ (= 3.11 fm), the projectile root mean square radius. ${ K}(r)$ is also seen to be constant for $r>8$ fm. These conditions have been checked to be true for the other three excited states of $^{16}$N.

\begin{table}[ht]
\begin{center}
\caption{Total one-neutron removal section ($\sigma_{-n}$) in the Coulomb breakup of $^{16}$N on Pb at 100 MeV/u, calculated at three different directions of local momentum for all the four low-lying states of $^{16}$N.}
\begin{tabular}{|c|c|c|c|c|}

\hline\hline
$J^{\pi}$  & Energy (MeV)   &  \multicolumn{3}{c|}{$\sigma_{-n}$ (mb)}  \\ 
\hline

 & & $d_1$ & $d_2$ & $d_3$  \\

\hline 
2$^-$ & 0     & 1.95  & 2.04  & 2.16\\
0$^-$ & 0.120 & 35.55 & 35.00 & 34.32\\
3$^-$ & 0.298 & 3.04  & 3.17  & 3.36\\
1$^-$ & 0.397 & 49.44 & 48.68 & 47.73 \\
\hline
\hline
\end{tabular}
\end{center}
\end{table}

In order to check the dependence of our results on the direction of ${\bf K}$, we calculate the total Coulomb breakup cross section at three different directions of the local momentum -- ($d_1$): parallel to the beam direction (zero angles), ($d_2$): parallel to the direction corresponding to the half of the angles of ${\bf q}_b$ and ($d_3$): parallel to ${\bf q}_b$. 

Table II, shows the variation of total cross section in the Coulomb breakup of $^{16}$N on Pb at 100 MeV/u, calculated at three different directions of local momentum as mentioned above, for all the four low-lying states of $^{16}$N. It is clear that the change in total cross section is less than 10 $\%$ for ground and second excited states and it is even less than 4 $\%$ for the first and third excited states, as one moves from direction ($d_1$) to ($d_3$).

\section*{Appendix B: Dependence on Woods-Saxon parameters}
\begin{figure}[h] 
\centering
\includegraphics[height=7.0cm, clip,width=0.45\textwidth]{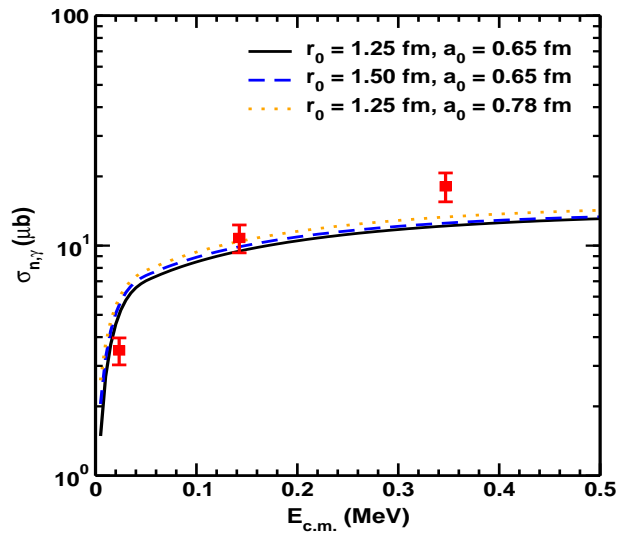}
\caption{\label{fig:8}(Color online) Variation of the total $^{15}$N$(n,\gamma)^{16}$N capture cross section with different Woods-Saxon parametrization. For more details see text.}
\end{figure}

We now investigate the dependence of the Woods-Saxon potential parameters on our results. In  
Fig. \ref{fig:8}, we show the variation of the total capture cross section for different combinations of the radius and diffuseness parameters which reproduce the same one neutron separation energy in $^{16}$N. The solid line shows the result that we have used in this paper. The dashed and dotted lines are those in which the radius and diffuseness parameters have been increased by $20\%$, respectively, over those shown by the solid line. We do not make out any major discernible difference in the results which could be validated by present day experiments.

\end{document}